\documentclass{emulateapj}
\def\eps@scaling{1.0}
\newcommand\plotthree[3]{%
 \centering 
 \leavevmode
 \columnwidth=.67\columnwidth
 \includegraphics[width={\eps@scaling\columnwidth}]{#1}%
 \hfil
 \includegraphics[width={\eps@scaling\columnwidth}]{#2}%
 \hfil
 \includegraphics[width={\eps@scaling\columnwidth}]{#3}%
}%

\def\eps@scaling{1.0}
\newcommand\myplottwo[2]{%
 \centering 
 \leavevmode
 \columnwidth=.97\columnwidth
 \includegraphics[width={\eps@scaling\columnwidth}]{#1}%
 \hfil
 \includegraphics[width={\eps@scaling\columnwidth}]{#2}%
}%

\begin{document}

\title{Gravitational Lensing of the Microwave 
Background by Galaxy Clusters}
\author{Gilbert Holder}
\affil{School of Natural Sciences, Institute for Advanced Study, Einstein Drive,
Princeton, NJ 08540}
\author{Arthur Kosowsky}
\affil{Department of Physics and Astronomy, Rutgers University, 
136 Frelinghuysen Road, Piscataway, NJ 08854-8019}

\begin{abstract}
Galaxy clusters will distort the pattern of temperature anisotropies
in the microwave background via gravitational lensing.  We create
lensed microwave background maps using clusters drawn from numerical
cosmological simulations. A distinctive dipole-like temperature
fluctuation pattern is formed aligned with the underlying microwave
temperature gradient.  For a massive cluster, the characteristic
angular size of the temperature distortion is a few arcminutes and the
characteristic amplitude a few micro-Kelvin. We demonstrate a simple
technique for estimating the lensing deflection induced by the
cluster; microwave background lensing measurements have the potential
to determine the mass distribution for some clusters with good
accuracy on angular scales up to a few arcminutes.  Future
high-resolution and high-sensitivity microwave background maps will
have the capability to detect lensing by clusters; we discuss various
systematic limitations on probing cluster masses using
this technique.
\end{abstract}

\keywords{cosmology: theory -- galaxies: clusters: general --
cosmic microwave background -- gravitational lensing}

\section{Introduction}
\label{sec:intro}

In the post-WMAP era of microwave background measurements, attention
is quickly shifting to smaller angular scales. At scales below 4
arcminutes, the temperature fluctuations are dominated not by
primordial fluctuations associated with the last scattering surface
but rather by secondary fluctuations induced by interactions with the
matter distribution at lower redshifts. Most prominent among these
effects are the Sunyaev-Zeldovich effect 
\citep{sunyaev72} and gravitational lensing.

This paper studies the gravitational lensing of the microwave
background by galaxy clusters. Much recent effort has been devoted to
lensing of the microwave background by the large-scale distribution of
matter \citep{oka03,hir03,kes03,oka02}, and
lensing of optical and radio sources by galaxies, galaxy clusters, and
large-scale structure is a major
theoretical and experimental industry. Curiously, little effort so far
has been put into lensing of the microwave background by clusters; the
only major work on the subject is the seminal paper by 
\citet{sel00}. 
The likely reason for this benign neglect is that only recently have
microwave maps of sufficient resolution and sensitivity to detect
cluster lensing become a realistic expectation \citep{kos03,woo02}.
The subject has recently been taken up by
\citet{dod03} as it relates to galaxies; \citet{cooray03} 
explicitly showed the dependence of the lensing signal on cosmological 
parameters, and \citet{bartelmann03} showed a map of a cluster lensing the CMB
as an application of numerical techniques for gravitational lensing. 

Observing cluster lensing of the cosmic microwave background requires
temperature variations on scales large compared to the cluster (so
that we have something to lens), but minimal variations on scales
comparable to the size of the cluster (so that the lensing signal can
be cleanly separated from the intrinsic temperature fluctuations).
Notably, this is exactly what the primary
microwave background fluctuations offer.  A
microwave background temperature map is dominated by fluctuations on
scales of a degree or larger, arising from density and temperature
perturbations at the surface of last scattering at a redshift $z\simeq
1100$.  At scales smaller than a degree, the power spectrum of
temperature fluctuations begins to decline due to diffusion damping:
perturbations on scales smaller than the thickness of the last
scattering surface are exponentially suppressed. On typical cluster
angular scales of a few arcminutes, the primordial temperature
perturbations are negligible. Secondary, non-linear temperature
fluctuations will arise, but these are generally either tiny
(e.g., gravitational lensing by large-scale structure) or have a
frequency dependence different from the blackbody distortions
due to lensing (e.g. the thermal
Sunyaev-Zeldovich effect). The only source of significant nearly-blackbody
fluctuations on scales of galaxy clusters besides lensing is the
kinematic Sunyaev-Zeldovich distortion, mainly due to the peculiar motion of
the cluster itself as well as a small component from
large scale structure \citep{vishniac87}. 
In general, the kSZ signal has a different
morphology than the lensing signal, and in most cases the difference
between the two will be clear. Furthermore, most of the kSZ signal should be
strongly correlated with the thermal SZ effect. However, the kSZ signal
will be a large source of contamination in the central regions, with a
significant source of noise coming from bulk motions within the cluster
due to objects that have recently been accreted.

Seljak and Zaldarriaga (2000) considered idealized spherical clusters
lensing a pure temperature gradient. They also provided an extensive
list of potential systematic effects which must be overcome to observe
the signal. In the space of three years, the sensitivity and
resolution of envisioned microwave background measurements have
increased dramatically, prompting a more detailed assessment of the
cluster lensing signal.  In this paper, we use model mass
distributions from clusters in a cosmological N-body simulation to
lens a background Gaussian temperature field constructed from the
temperature power spectrum in a realistic cosmological model.  The
cluster lensing signal is clear in the resulting map. Section II
displays how the signal varies with cluster mass
and cluster location on the sky, and Section III considers how well
the cluster lensing signal can be inferred from an ideal map using 
a correlation function technique for estimating the unlensed temperature
distribution. More elaborate
model-fitting techniques are likely straightforward but beyond the
scope of the paper, requiring explicit and careful considerations of
specific instruments and observing strategies. 
Finally, we discuss potential advantages and
systematic limitations of this method for cluster mass determination
in the context of realistic experiments.

\section{Order-of-Magnitude Estimates}

The root-mean-square temperature gradient in the primary microwave 
background fluctuations
is around 12 $\mu$K per arcminute for standard cosmological models
consistent with the measured microwave power spectrum; the magnitude
of the lensing signal is on the order of the local temperature
gradient times the characteristic angular deflection. Typical
deflection angles (roughly the cluster gravitational
potential in units of $c^2$, $\Phi/c^2\simeq 10^{-4}$) 
are on the order of tens of arcseconds. The
fractional perturbation due to lensing is therefore on the order of
a few $\mu$K.

This is at least an order of magnitude larger than perturbations due
to the transverse velocity of the cluster \citep{birkinshaw83}, 
which is on the order of the cluster potential $\Phi/c^2$ times
the transverse velocity
$v_{trans}/c$. Typical peculiar velocities are a few hundred km/s, 
so this term should be at the level of roughly 0.1 $\mu$K. 
Similarly, anisotropy
imprinted from the cluster potential changing over the photon
travel time (the Integrated Sachs-Wolfe effect in the non-linear regime) 
should be roughly given by the potential times the
photon travel time relative to the dynamical time. With travel times
on the order of 1 Myr and dynamical times on the order of 
1 Gyr this term should be comparable to the transverse velocity 
effects but well below the expected level of lensing effects.
Therefore, the signatures of moving or dynamically evolving clusters
will be strongly suppressed compared to lensing of the CMB, 
even if the thermal
and kinematic SZ effects can be efficiently removed.

As discussed below, the kinematic SZ effect is also expected to
be on the order of a few $\mu$K, with a similar spectral dependence
to gravitational lensing. These two signals must be separated via
their spatial morphology or other techniques.

\section{Lensed Maps}

\begin{figure}
\plotone{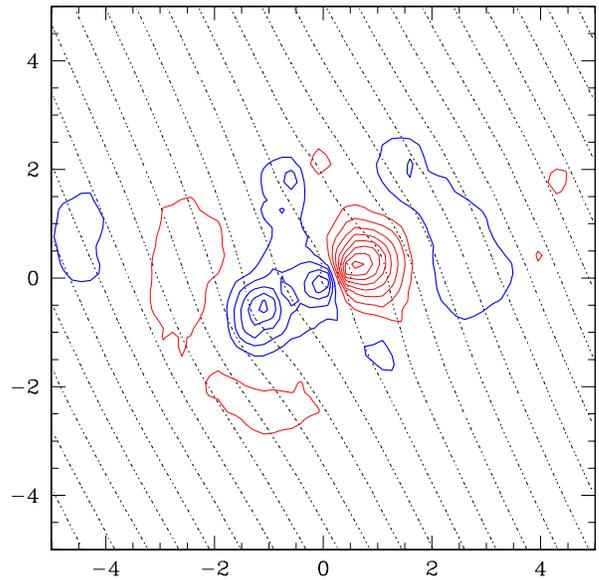}
\caption{Gravitational lensing of a pure background gradient of
15 $\mu K$ per arcminute by a galaxy
cluster of mass $7\times 10^{14} h^{-1}M\odot$ 
extracted from a numerical simulation. Background contours
are spaced by 5 $\mu K$ and show the unfiltered lensed signal, while
solid contours, spaced by 0.5 $\mu K$, show maps that have been high-pass 
filtered, with all $k\lesssim 1150$ ($\ell \lesssim 7200$)
removed. Each image is $5'\times 5'$ and does not include 
kinematic SZ effects. Blue (thick) contours indicate positive 
temperature differences, while red (thin) contours indicate negative values.}
\label{fig:grad}
\end{figure}

\begin{figure}[t]
\plottwo{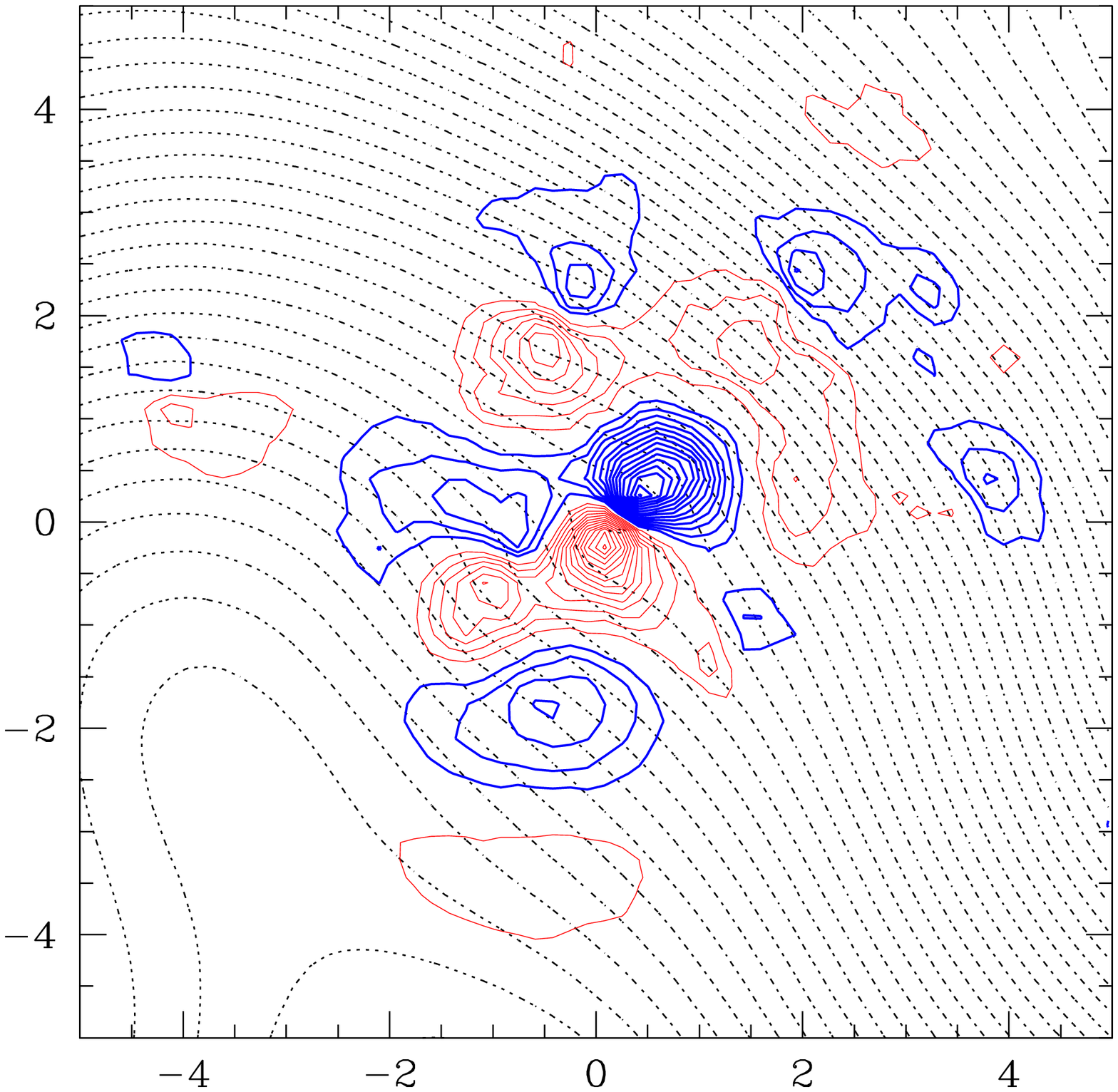}{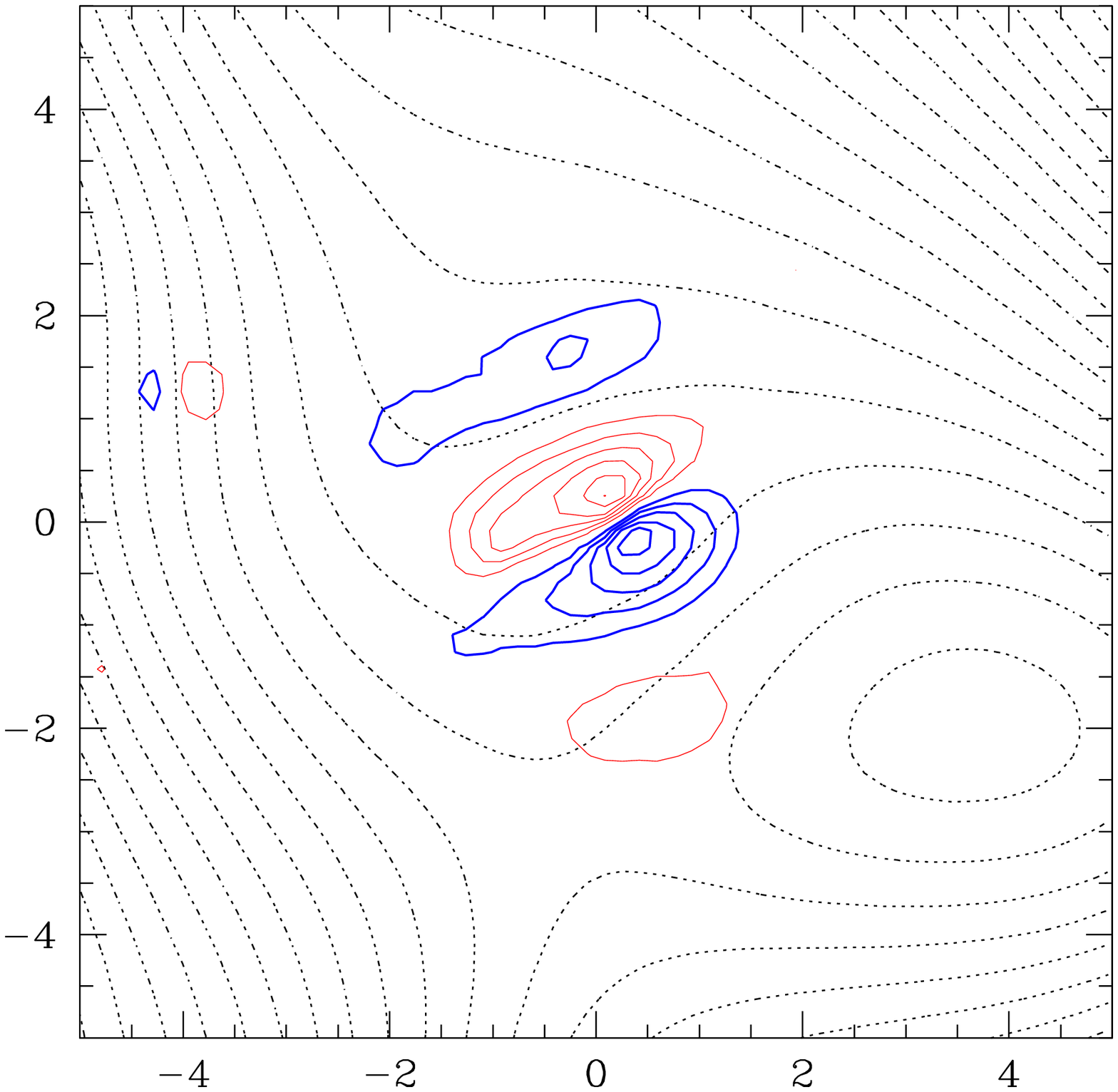}
\caption{Same as Fig.~1, except that the cluster
is lensing two different realizations of a CMB sky.}
\label{fig:cmb}
\end{figure}

Figure \ref{fig:grad} displays the lensing of a pure temperature
gradient by a numerically simulated cluster of galaxies of mass
$M=6h^{-1}\times 10^{14}M_\sun$ at a redshift of $z=0.5$, while 
Fig.~\ref{fig:cmb}
shows the same cluster lensing two realizations
of a Gaussian random temperature field with the power spectrum of the
microwave background for a flat $\Lambda$CDM cosmological model with
$n=1$, $h=0.7$, $\Omega_b h^2 = 0.024$, and $\Omega_\Lambda=0.7$.
Equal-temperature contours are plotted at 5 $\mu$K separations as
light dotted lines.  To display the lensing signal more clearly, we
then high-pass filter the map with a 3-arcminute filter scale.  The
temperature contours of the filtered map are plotted at 0.5 $\mu K$
separations as heavy lines.  

The cluster is drawn from the
VIRGO\footnote{http://virgo.sussex.ac.uk} simulations \citep{sil00}.
Outputs (including gas) were available at $z=0$, and for our lensing
calculation we artificially placed the simulation volume at $z=0.5$.
To be conservative, 
we did not scale the box size to account for the expansion of the
universe between $z=0.5$ and $z=0$.  
This scaling would lead to an overestimate of the central
concentration of virialized objects at $z=0.5$ in the simulation, 
and therefore an overestimate of the lensing signature.  
The box was translated such
that a massive cluster was at the center of the projected mass
distribution, and the dark matter distribution was projected onto a
$2048\times 2048$ grid using a simple nearest grid point method. The
gas was smoothed (in projection) over the mean distance to the 24th
nearest neighbor assuming a uniform disk smoothing kernel.  The
resulting surface mass density map was used to generate deflection
maps with resolution of roughly 10'', using an FFT method.

We compute the reduced angular deflection vector due
to the cluster mass distribution in the thin-lens approximation, 
\begin{equation}
\mathbf{\alpha}({\bf x}) = {1\over \pi}\int \kappa({\bf x'})
{{{\bf x}-{\bf x'}}\over{\left|{\bf x}-{\bf x'}\right|^2}} d{\bf x'}
\label{displacement}
\end{equation}
where ${\bf x}$ is a 2-dimensional vector describing the angular
position on the sky, $\kappa({\bf x})$ is the projected
convergence related to the surface mass density $\Sigma$ by
\begin{equation}
\kappa({\bf x}) = {4\pi G\over c^2}{D_{\rm d}D_{\rm ds}\over 
D_{\rm s}} \Sigma({\bf x})
\label{convergence}
\end{equation}
with $D_{\rm d}$, $D_{\rm s}$, and $D_{\rm ds}$ the
angular diameter distances from the observer to the lens,
the observer to the source, and the lens to the source, respectively.
The actual deflection vector ${\bf\hat\alpha}$ 
is related to the reduced deflection by the distance scaling 
${\bf \alpha} = (D_{\rm ds}/D_{\rm s}){\bf \hat\alpha}$.
(See, e.g., \cite{sch92}
for a detailed exposition of standard gravitational lensing
theory.)  

A lensed map can be constructed from an unlensed background map and
a surface mass distribution $\Sigma({\bf x})$ by 
solving the lens equation
\begin{equation}
\mathbf{\beta}  ({\bf x}) = {\bf x} - \mathbf{\alpha}({\bf x})
\label{lens_equation}
\end{equation}
where ${\bf\beta}$ is the source position in the image plane.
For a source covering the entire sky like the microwave background, 
the lens equation is straightforward to solve
by pixellating the image plane, then using the displacement vector
to map each point in the image plane to a point of the source
plane, then assign the source temperature at that point to the
original point in the image plane.
The resulting lensed maps in Fig.~\ref{fig:grad} 
represent the signal an ideal
experiment with very high angular resolution and sensitivity
might measure. 

The CMB maps were generated using a CMB power spectrum generated
by CMBFast\footnote{www.cmbfast.org} \citep{seljak96}. To be conservative,
we use
the lensed power spectrum, which has slightly more power on small scales,
but assume the underlying CMB anisotropies are a Gaussian random field.
We construct unlensed maps over small regions of sky 
using the flat sky approximation.

The cluster lensing signals are obvious in the filtered maps. It is
clear that lensing induces structure in the maps on significantly
smaller scales than those on which the microwave background has
significant temperature fluctuations.  For a cluster
situated in front of a region of the microwave background which is
approximately a temperature gradient, lensing produces a
characteristic dipole-like pattern, with a cool and a hot peak. The
peak-to-peak amplitude is proportional to the gradient magnitude, but
is generally on the order of 1 to 10 $\mu$K; the angular separation of
the peaks is on the order of an arcminute, with noticeable lensing
effects out to radii of several arcminutes. Also note that a vector
from the hot to the cool lensing peak must be in the direction of the
local CMB gradient: this distinctive signature can be used to
discriminate between a lensing signal and other effects local to the
galaxy cluster, which will be physically uncorrelated with the
microwave background.

The gross features of the cluster lensing signal depend on cluster
mass, redshift, and sky location. The lensing effect depends on the
angular diameter distance to the cluster; for massive clusters which
are distant enough that their angular size is on the order of an
arcminute, their angular diameter distance is only a weak function of
redshift. The lensing deflection depends primarily on the total mass
of the cluster.  The other crucial factor in the cluster lensing
pattern is sky position: the amplitude of the lensing signal is
proportional to the local background temperature gradient.  To extract
mass information about the cluster, we must be able to infer the
unlensed background temperature pattern. This is easiest when the
cluster is situated in a sky region which has a fairly uniform
temperature gradient; in this case the characteristic dipole lensing
pattern is produced. If the cluster is located in a region of the sky
where the temperature field is near an extremum, or where the
isotemperature contours are strongly curved, extraction of the lensing
mass is more uncertain, because the unlensed temperature field is not
as well constrained by the lensed map.

\begin{figure}
\plotone{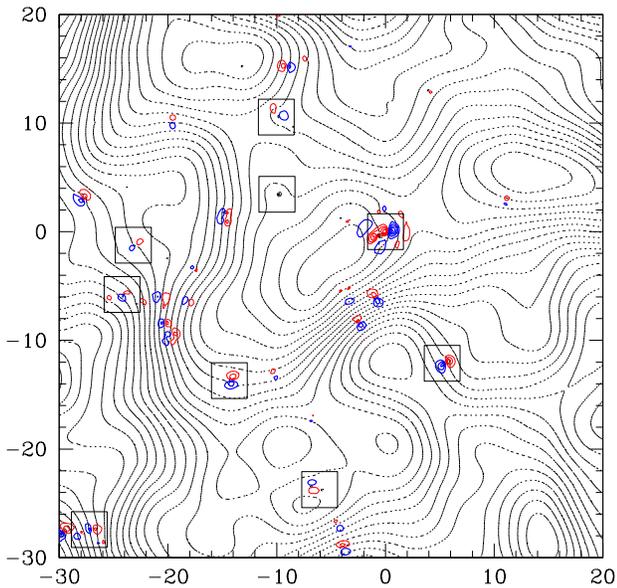}
\caption{$50' \times 50'$ 
field showing several clusters lensing the CMB.
All positions of clusters with masses above 
$10^{14} h^{-1}M_\odot$ are shown as open squares in the map, and
background dotted contours are spaced every 15 $\mu K$. Solid contours
show a high-pass filtered map with 0.5 $\mu K$ temperature
contours, as in Fig.~\ref{fig:grad}.}
\label{fig:bigger}
\end{figure} 

Figure \ref{fig:bigger} shows clusters with a range of different
masses lensing the microwave background. Here the simulated lensing
mass distribution is a cubic volume $100h^{-1}$ Mpc on a side drawn
from the VIRGO simulations, placed at a redshift of $z=0.5$.  Both the
amplitude of the lensing peaks and their angular separation increase
with cluster mass. To gauge the effects of sky position, we have
outlined with a square the location of all clusters in the simulated
mass distribution with $M>10^{14}\,M_\odot$.  Some marked clusters do
not obviously correspond to clean features in the filtered map,
indicating that the lensing signature is very difficult to observe at
these positions. Roughly half of the $10^{14}\, M_\odot$ clusters
in Fig.~\ref{fig:bigger} are in sky positions leading to appreciable
lensing signals. As a rule of thumb, the lensing signal will be
easiest to estimate when $|\nabla T|$ is large and $\nabla^2 T$ is
small. At the same time, occasional less massive objects that are
particularly well-placed at the positions of strong gradients display
clear lensing signatures, as with the cluster near $(-30',5')$.
Note that the positions of all clusters will be known from their
thermal Sunyaev-Zeldovich distortions, which are at least an order of
magnitude larger than the lensing signature. 

\begin{figure}
\plotone{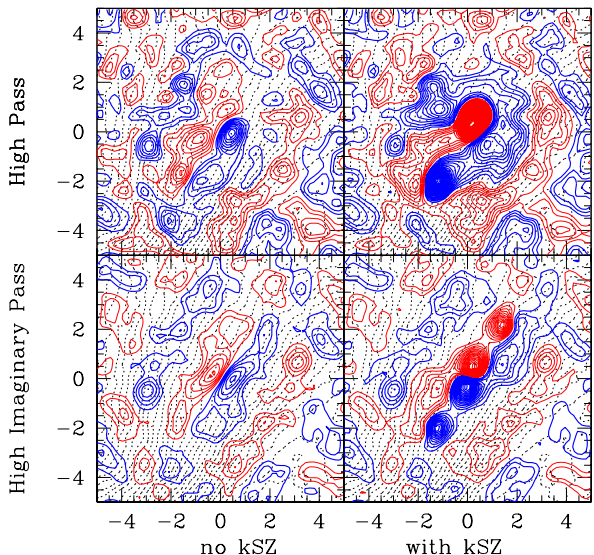}
\caption{Effects of kinetic SZ signal: left two panels show case of no
kSZ, right two panels include a large contaminating kSZ signal. Top panels
show a simple high-pass filter, as in figure 1, while bottom two panels
show the effect of a high-pass filter that has also selected the 
component that is anti-symmetric around the cluster center (i.e., the
imaginary part of the Fourier transform).}
\label{fig:ksz}
\end{figure}

In Figure \ref{fig:ksz} we show the problems that can be introduced by 
significant
substructure in the kSZ signal. This case is chosen as an unfavorable
kSZ signal, although it could be even less favorable if aligned exactly with
the gradient. Some of the kSZ is removed by selecting only that component
that is anti-symmetric around the cluster center (as the lensing signal
should be). Along the direction of the gradient, it is still possible to
see the characteristic lensing distortion, but it is clear that the central
region will be useless for the purposes of lensing reconstructions. We
emphasize that this was not selected as a typical contamination, and the
simulation physics (no feedback or preheating) led to gas cores that are
much more concentrated than in observed clusters.

\section{Deflection Estimation}

The maps in the previous section make clear that in many cases (but
not all), some estimate of cluster mass within a certain radius can be
made from the pattern of lensing in the microwave background.  It is
likely that some kind of model fitting for both the cluster mass
profile and the unlensed temperature map will give the most accurate
cluster mass recovery, but a complete parametric study is beyond the
scope of this paper. Here we present a crude technique for estimating the
lensing deflection in the inner region of the cluster which works well in
many cases. This deflection angle can then be used as a constraint on
the mass distribution.

\subsection{Wiener Estimation of Unlensed Background}

For the case of a source intensity distribution which is a pure
gradient, it is straightforward to estimate the lensing deflection in
the direction along the temperature gradient. At sufficiently large
distances from the cluster center, the background gradient will be
clear and can be robustly estimated. For the case of the microwave
background, a pure gradient is generally {\it not} a good description
of the background temperature distribution on scales of a few
arcminutes over which lensing deflections are significant.  However,
the background temperature distribution is close to a gaussian random
field, and the statistical nature of this distribution can be used to
estimate the unlensed background temperature distribution in the
region of a cluster: we mask out the central regions of the image,
where lensing is known to be important, and interpolate an approximate
unlensed temperature distribution using the region that is more than a
few arcminutes away from the center.

The most reliable method we found is Wiener interpolation.
We assume that the correlation function of the 
microwave background temperature
is known from power spectrum measurements.
Then we have a good estimate of $\left\langle T_i
T_j\right\rangle = C(|{\bf\theta}_i - {\bf\theta}_j|) \equiv C_{ij}$, 
the mean product of temperatures at two points ${\bf\theta}_i$ and
${\bf\theta}_j$ in the
map; here $C(\theta)$ is the temperature correlation function for
two points with angular separation $\theta$. The matrix ${\bf C}$ is
just the theoretical gaussian random 
correlation matrix constructed from the power spectrum. 
Given accurate estimates of the temperature map outside the cluster
region, we can use these correlation functions to interpolate across
the cluster.

The pixels of some observed map can be viewed 
as a single data vector ${\bf d}$. We then form the
noise-weighted covariance matrix ${\bf D}$: this covariance
matrix includes the theoretical covariance ${\bf C}$ plus
the covariance arising from pixel noise. 
We mask out given pixels in the region
of a cluster by setting the diagonal elements of ${\bf D}$
corresponding to those pixels to a value which is
large compared to the other elements of ${\bf D}$. 
We then estimate the temperature map ${\bf t_{est}}$
defined by the theoretical correlation function which matches
the measured pixels in the region ouside the cluster as
${\bf t_{est}} = {\bf C}{\bf D}^{-1}{\bf d}$. 
This construction requires inversion of a large
matrix.  For a map with $N \times N$ pixels, the computation of the
inverse covariance matrix scales as $N^6$, but 
pixels well beyond the cluster region do
not contribute much to the cluster region. Also, note that this
inversion must be done only once for a given microwave background
power spectrum, pixel scale, and mask geometry.

\begin{figure}
\plotone{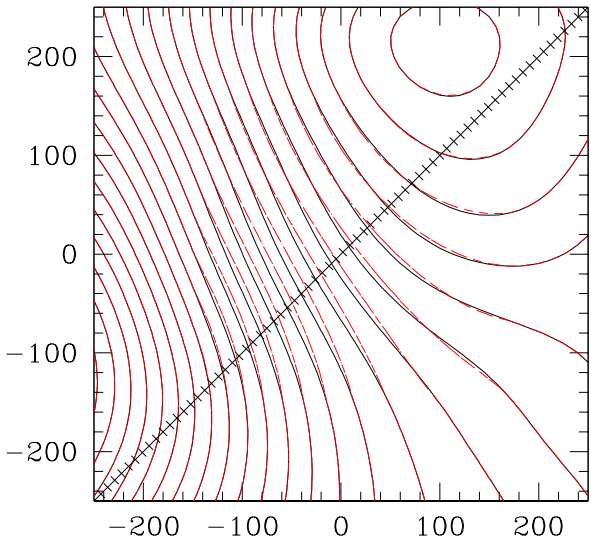}
\caption{Reconstruction of unlensed CMB background field using Wiener
interpolation. The axes are in arcseconds, and the diagonal line is marked
in units of 5 arcseconds; temperature contours are spaced at 5 $\mu$K
intervals. Reconstruction of the central $5'\times 5'$
section of the map is done using the displayed
map region with an effective pixel scale of 10''. Solid lines show the
input map, dashed lines show the reconstructed map.} 
\label{fig:nolens_recon}
\end{figure}

As a verification of the method, we start with a CMB
realization (with no cluster lensing), 
cut out the central 8' and use the Wiener method to
reconstruct the missing pixels. Results are shown in Figure
\ref{fig:nolens_recon}.  The reconstruction is not perfect, but the
reconstruction errors are on the order of 5'', small compared to
typical lensing deflection angles of 30-40''. For
Fig.~\ref{fig:nolens_recon}, we used a map that was $64\times 64$
pixels with a 10'' pixel scale. This reconstruction took a few minutes
on a desktop computer without any particular optimization effort.

We also experimented with fitting a simple gradient or two-dimensional
splines of varying stiffnesses, with varying degrees of success on a
case-by-case basis. For the Wiener interpolation technique, assuming
the outer regions are unaffected by lensing introduces an unnatural
boundary condition, so an iterative approach would improve this
estimate. Ultimately, a simultaneous fit for the unlensed CMB
distribution, kinematic SZ, and deflection angle will be required for
the most reliable constraints, and will use the corresponding thermal
SZ map to provide a rough template for the kinematic SZ signal.

We note that the Wiener estimation technique provides a general method 
for 
testing Gaussianity of the microwave background temperature distribution.

\subsection{Deflection Angle Reconstruction}

\begin{figure}
\plottwo{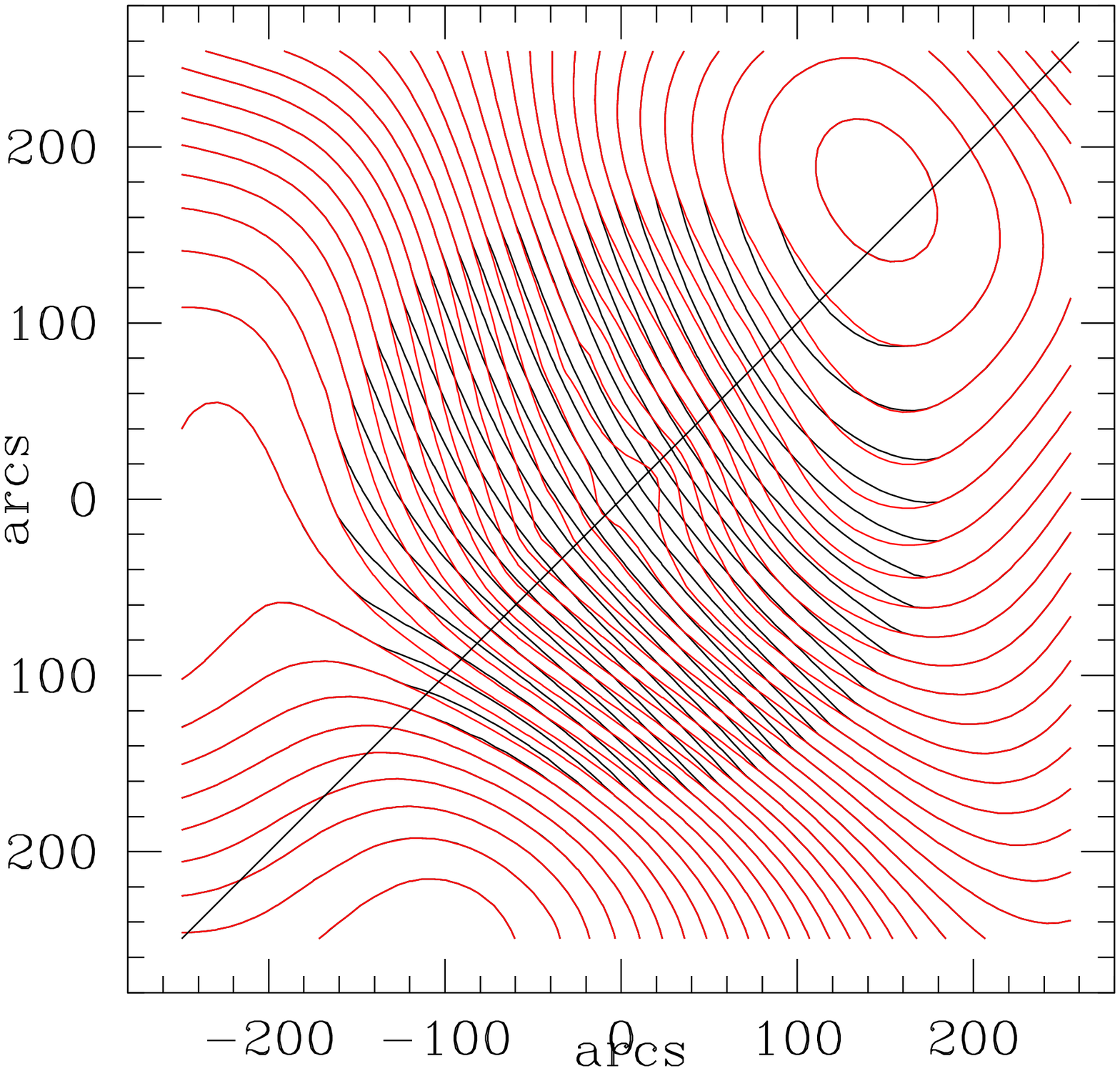}{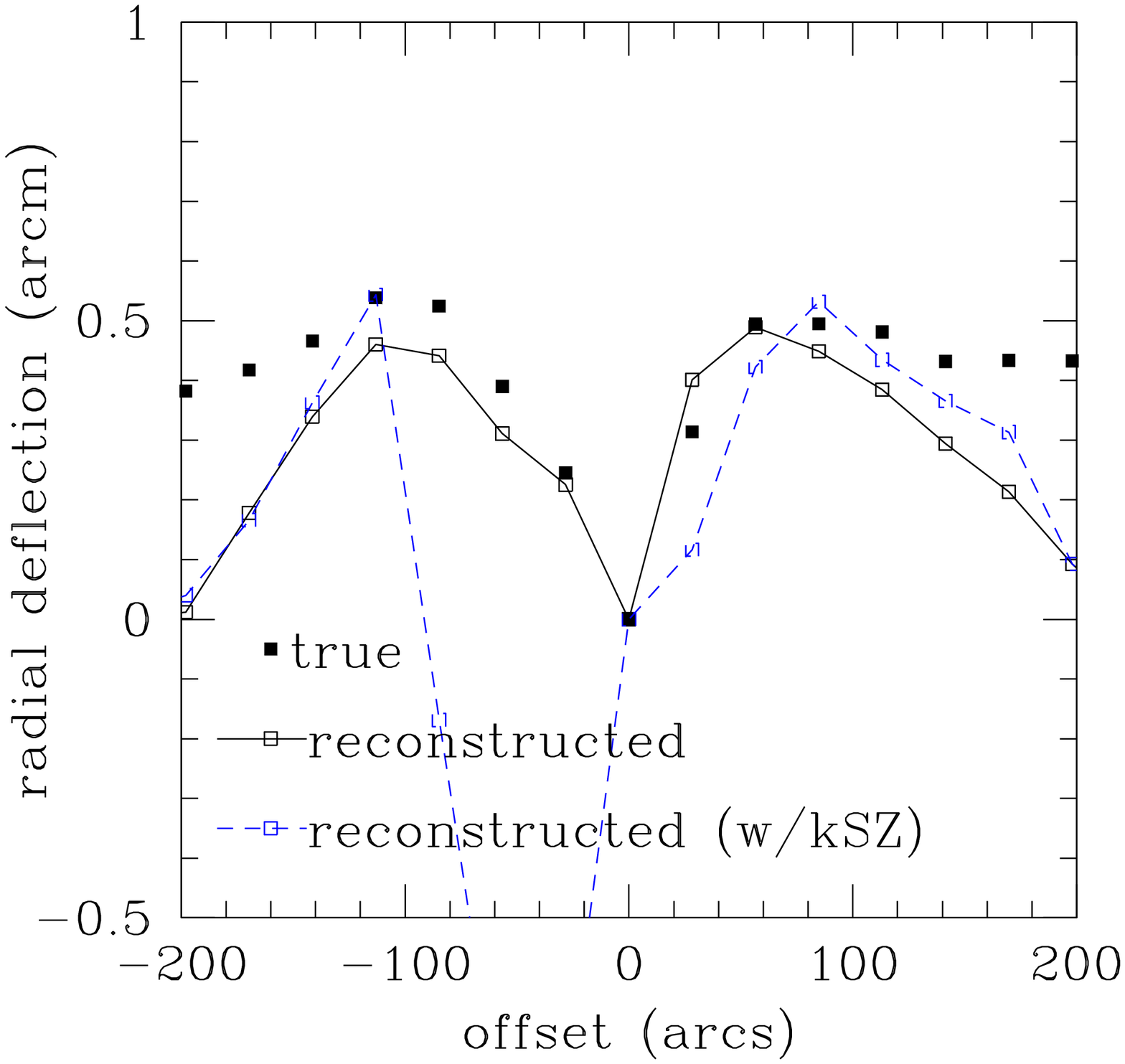}
\caption{Reconstruction of lensing deflection angle. Left panel shows
lensed and reconstructed unlensed CMB map; the axes are
in arcseconds. The diagonal line shows the
slice that was used to reconstruct the radial deflection angle shown
in the right panel. The solid curve shows the reconstruction in the
case where the kSZ can be perfectly removed, while the dashed line
shows the deflection estimate when no kSZ has been removed. }
\label{fig:lens_recon}
\end{figure}

Applying the Wiener estimation method, we show the reconstructed
deflection angle along a slice through the cluster center in 
Fig.~\ref{fig:lens_recon}. At the edge of the masked region, the
deflection goes to zero by construction; it may be useful to employ
an iterative approach to more accurately reconstruct the outer regions.
The inner few arcminutes are accurately reconstructed, providing 
information on the cluster mass distribution.

A cluster will produce a second microwave background distortion with a
near-blackbody spectrum in addition to lensing, namely the kinematic
Sunyaev-Zeldovich (kSZ) effect \citep{sunyaev72}. This distortion 
arises from
photons scattered by the cluster gas moving with
some bulk velocity relative to the microwave background rest frame.
If clusters were balls of gas with no significant internal motions, the
result would be either a positive or negative temperature distortion
proportional to the gas density profile, for clusters with radial
velocities toward or away from the observer. This is easily
distinguishable from the dipole-like pattern induced by
lensing. Furthermore, this spatial pattern would be matched by the thermal
SZ signature, and could be projected out. 
However, actual clusters possess significant internal
coherent bulk velocities, particularly for higher-redshift clusters
undergoing frequent mergers, and these internal velocities result in
more complex kSZ distortions (see, e.g., \cite{nag03}).
Occasionally, the resulting kSZ maps can even mimic the dipole-like
pattern seen in lensing, although alignment with the temperature
gradient would be purely coincidental.

The degree to which the kSZ effect will degrade mass estimates will
vary between individual clusters. Generally, the region of significant
kSZ distortion is more spatially concentrated than the lensing signal,
and will have some correlation with the thermal SZ distortion since
both scale with the cluster's total optical depth.  
Figure \ref{fig:lens_recon} also shows
recovered deflection angles when the kSZ effect is included.
Clearly, the lensing reconstruction is seriously degraded in
the central cluster region where the kSZ signal is comparable
to the lensing signal, but for regions further from the center,
the kSZ signal does not alter the reconstruction. In practice,
the temperature fluctuations in an annulus from 1' to 3' away from
the cluster will provide most of the information about lensing
deflections.

We have not included residual noise from large scale structure
\citep{vishniac87}, which should be at the level of roughly
1 $\mu$K at the resolution of these maps. While this noise makes
maps less visually appealing at the lowest contour levels, it does
not quantitatively affect the ability to reconstruct deflection angles.
Assuming typical gradients of 12 $\mu$K per arcminute, noise at the
level of 1 $\mu$K corresponds to fluctuations at the level of 5'' in
the deflection angle.

All of the above analysis has been for idealized maps, which
here is equivalent to an angular resolution of $10''$ and
noise levels below 1 $\mu$K per resolution element. The next
generation of microwave background maps at small scales
(e.g., Atacama Cosmology Telescope; ACT \cite{kos03} or 
South Pole Telescope; SPT) will attain resolutions of
roughly 1' and noise levels of a few $\mu$K per resolution element.
The ability to detect actual lensing deflections at these
resolutions and sensitivities will be lower than presented here;
quantitative analysis of how much the signal is degraded is
ongoing. Eventually, the Atacama Large Millimeter Array, 
ALMA\footnote{www.alma.nrao.edu} \citep{woo02}, 
will have both the sensitivity
and resolution to make the kinds of maps displayed in this paper,
provided systematic errors can be controlled well enough to allow reliable
imaging of more than ten square arcminutes (requiring a large mosaic).

\subsection{Mass From Deflection}

In an ideal case, lensing of the microwave background determines the
component of the deflection angle in the direction of the local
temperature gradient. In principle, this is not sufficient information
to determine the mass distribution. The deflection ${\bf\alpha}({\bf
x})$ can be expressed as the gradient of a lensing potential,
${\bf\alpha} = \nabla\psi$; thus the deflection field is curl-free,
$\nabla\times{\bf \alpha} = 0$. Thus if we infer $\alpha_x$ from a map,
in principle this determines $\partial\alpha_y/\partial x$ as well
(though of course no practical algorithm should differentiate measured
data!).  However, we cannot obtain $\alpha_y$ itself by integrating
with respect to $x$, because a different integration constant obtains
for each value of $y$. So we cannot recover the complete deflection
angle ${\bf\alpha}$ nor the convergence (mass density) 
$2\kappa = \nabla \cdot\ {\bf \alpha}$, but rather one component
of the deflection and the derivative of its orthogonal
component.

In practice, a number of other reasonable assumptions or
constraints can be imposed. Assuming the deflection to
be zero beyond some given radius will break the degeneracy
while biasing the cluster mass estimate. 
The assumption of a spherically symmetric lens will
also break the degeneracy; more generally, assuming almost
any parametric form for the lens will largely lift the
degeneracy.

\subsection{Errors in Cluster Catalogs}

So far we have discussed galaxy clusters individually. Future 
high-resolution microwave maps will yield samples of hundreds
to thousands of galaxy clusters selected by their thermal
SZ spectral distortions. The lensing signals in these samples can
be used to probe the cluster mass distribution in a statistical
sense. In this case, it is important to distinguish statistical 
errors, which will average out in a large cluster sample, from
systematic ones which will not.

The major sources of error affecting the lensing 
observations and analyses described here are residual uncertainties
from projecting out the thermal SZ spectral distortions and other
non-blackbody foreground components like dust or point sources; separation
of the kSZ signal from the lensing plus background CMB signals; 
errors in reconstructing the deflection map from the lensed CMB
map; and other neglected sources of signal like the moving cluster
effect. In order for an error to have a systematic impact on
lensing mass estimates, it must correlate with the cluster
lensing signal, which is aligned with the local temperature
gradient. This means that no foreground emission is likely to
contribute any systematic error to the lensing signal. The separation
of the thermal SZ signal is also unlikely to give any systematic
error, because of its roughly symmetric shape. The moving cluster
effect is both negligibly small for any single cluster and also
uncorrelated with the background temperature gradient so it will not
give any systematic effect over a significant survey area (i.e., a few
square degrees). The kinematic SZ effect can coincidentally
mimic the lensing signal to some extent in the case of a cluster
undergoing a merger, but the dipole-like morphology of the signal
will only be aligned with the background temperature gradient by 
coincidence in a small fraction of these merging clusters; also
any kinematic SZ dipole-shaped signature will be over smaller angular
scales than the lensing signal and will mostly be eliminated by masking
out the central portion of the cluster.

The only clear potential systematic error arises from the technique
used to separate the lensing signal and the background fluctuations.
In our sample technique in this paper, for example, we reconstruct
the deflection over some finite-area region around the cluster,
assuming zero deflection at the edges of the region. This leads
to a systematic underestimate of the deflection in the outer
parts of the cluster, as seen in Fig.~\ref{fig:lens_recon}. The
extent to which this systematic error biases ultimate mass estimates
depends on the details of the method used to reconstruct the deflection.
This bias depends primarily on the shape of the background fluctuations,
and for a given reconstruction technique can probably be modelled
with a fair degree of certainty. Detailed estimates of the size of
this systematic error when constructing statistical estimates from
cluster catalogs is beyond the scope of this paper. 

\section{Discussion}

Cluster mass estimates based on the gas distribution, such as those
extracted from X-ray temperature maps or from the Sunyaev-Zeldovich
effect, have intrinsic uncertainties arising from the detailed
relationship between the gas distribution and the total mass
distribution, particularly at higher redshifts when clusters undergo
frequent mergers. The only direct way to measure cluster masses is
through gravitational lensing. Strong and weak lensing of background
galaxies by a cluster give measurements of the shear field induced by
the cluster mass distribution, and elaborate techniques have been
developed to reconstruct an estimated mass distribution from lensing
observations \citep{kaiser93,squires96}. 
A number of cluster masses have been measured
this way \citep{fis97,tys98,wit01,kin02}.

However, cluster mass determination via lensing of background galaxies
has a number of limitations. At usual magnitude limits,
most clusters do not
exhibit the arcs of strongly lensed background galaxies, which arise
due to chance alignment of the cluster and background galaxy.  Mass
measurements for large samples of galaxy clusters thus cannot rely on
strong lensing as a primary tool. Weak lensing of faint background
galaxies provides a more generic signal. But weak lensing has a number
of difficult systematic considerations: (1) Orientations of background
galaxies possess some degree of intrinsic alignment, which can bias
the lensing signal \citep{hea00,pen00,cro01,cat01,cri01,bro02}.  
(2) The redshift distribution
of background galaxies is generally not well determined, and
photometric redshift distributions may induce 
significant systematic errors
\citep{smith01,mckay02,sheldon03}. 
(3) The lensing signal becomes weaker for higher
cluster redshifts because fewer background galaxies are behind
high-redshift clusters. (4) At a given exposure depth, 
weak lensing measurements are limited by
shot noise due to a finite number of background galaxies.  

Lensing of the microwave background ameliorates these limitations.
The photons all originate from a single, precisely-determined redshift.
and this surface of last scattering is well behind any
cluster, simplifying the redshift dependence of the lensing signal 
Furthermore, given sufficient resolution, the lensed temperature
pattern can be probed with arbitrary accuracy, circumventing the shot
noise limits. An additional potential advantage, pointed out by Seljak
and Zaldarriaga (2000), is that deflection angle drops off more
slowly with distance from the cluster than does shear. However,
as radius increases, the comparatively larger lensing signal is
counteracted by the larger noise (arising from the
primary CMB anisotropies) from uncertain knowledge of the
unlensed background; in practice this advantage over weak
lensing is uncertain.

The microwave background lensing signal will have its own systematic
limitations. Foremost among these are the contribution of the
kSZ signal and our limited knowledge about the underlying primordial
temperature fluctuation pattern. As illustrated here,
the impact of each of these varies from cluster to cluster. It is
safe to say that some non-negligible fraction of clusters will be
good candidates for lensing mass determinations, due to their
location in front of a favorable area of temperature fluctuations 
and their lack
of significant internal bulk velocities producing a complicated
kSZ contribution. Also, which clusters will provide the most reliable
lensing deflection determination will be evident from the maps
themselves.

Finally, any lensing mass estimate does not probe only the cluster
mass, but rather all of the mass projected along the line of sight to
the cluster. Generally, this total projected mass is dominated by the
galaxy cluster mass, but other masses will lead to a small systematic
bias towards overestimating cluster masses. This can be addressed by
comparing observations to full cosmological simulations instead of to
a simple cluster mass distribution \citep{chen03}. 
Note that weak and strong lensing
of background galaxies also suffer from this difficulty, although
the larger redshift of the microwave background makes the effect
somewhat more important. The linear contribution due to large scale
structure has been crudely included by our use of the lensed CMB
power spectrum, but this does not adequately include the effects
of non-Gaussianity and non-linearity.

Of course, the usual slew of difficulties in making accurate,
high-resolution and high-sensitivity microwave background maps will
also need to be overcome, including foreground emission, point
sources, and the cluster's comparatively large thermal
Sunyaev-Zeldovich distortion.  Confusion-limited infrared point
sources, in particular, will be a serious problem at arcminute
resolutions \citep{knox03,borys03,blain98}.  The cluster thermal SZ
effect can be minimized by observing near the frequency null around
220 GHz. Thermal SZ and foreground separation at the $\mu$K level
will require multi-frequency observations, 
although both will be uncorrelated with
the specific lensing temperature distortions around clusters. Accurate
measurement of lensing signals with maximum amplitudes of a few $\mu$K
in individual clusters will likely require sub-arcminute resolution
observations over a range of frequencies by an instrument like ALMA, and
may be marginally detectable with ACT or SPT.

Lensing of the CMB polarization fluctuations should be less susceptible
to many sources of confusion than lensing of the temperature
fluctuations considered in this paper: the kSZ and thermal SZ polarization
fluctuations are significantly smaller than the corresponding
temperature fluctuations, and infrared point sources, which are
actually the total dust emission from high-redshift galaxies, are
likely to show little polarized emission.  Furthermore, the primary
CMB polarization pattern on small scales will have a particular spatial
property (curl-free) that is violated in the presence of lensing
\citep{zal97,kam97},
providing a ``smoking gun'' of the presence of lensing distortions. If
the systematic limitations to measuring lensing deflections from
temperature maps prove difficult to overcome, polarization maps
provide a viable alternative (although they require still greater
experimental sensitivity due to the smaller polarization amplitude in
the primary microwave background fluctuations).

The importance of constraining galaxy cluster masses hardly needs to be
emphasized. The number density of clusters of a given mass as a
function of redshift is a highly sensitive probe of the growth rate of
structure, and in principle could strongly constrain the scale factor
evolution at recent cosmological epochs \citep{haiman00}.  The
problem is that traditional cluster observations in optical and X-ray
bands, or measurements of the Sunyaev-Zeldovich distortion, measure
the distribution of baryons in the cluster, not its total
mass. Connecting the distribution of baryons and dark matter in
clusters is a difficult and complex problem, providing many
possibilities for systematic errors in cluster mass estimates,
particularly at larger redshifts where clusters are further from
dynamical equilibrium due to frequent merging. Lensing measurements
appear to be the only route to a direct, reliable cluster mass
estimate. A clear understanding and characterization
of cluster lensing is also essential for probing the cluster
kinematic SZ signal \citep{hol03,nag03}, where the lensing acts as 
an unwanted complication. We hope that the images and
arguments in this paper convincingly demonstrate that microwave
background lensing is a viable alternative to weak lensing of
background galaxies for learning about the distribution of mass in
galaxy clusters. 

{\em Note}: since submission of this work, two noteworthy papers of relevance
have appeared in the literature \citep{vale04, dodelson04}. These works
confirm the results presented here and present interesting discussions
of other important issues and directions for lensing of the CMB by 
galaxy clusters.

\acknowledgements
This work has been partially supported by NASA 
Space Astrophysics Research and Analysis grant NAG5-10110 at Rutgers. 
A.K. is a Cottrell Scholar of the Research Corporation, and G.H. is
supported by the W.M. Keck Foundation.

\end{document}